\documentclass[aip,cha,reprint]{revtex4-1}
\usepackage{amscd}
\usepackage{amssymb}
\usepackage{amsfonts}
\usepackage{amsmath}
\usepackage{graphicx}
\usepackage{dcolumn}
\usepackage{bm}
\usepackage{multirow}
\usepackage{hyperref}

\begin{document}

\title{Multiscale characterization of recurrence-based phase space networks
constructed from time series}

\author{Ruoxi Xiang}
\email{rxxiang@gmail.com}
\affiliation{Department of Electronic and Information Engineering,
Hong Kong Polytechnic University, Hong Kong, People's Republic of
China}
\author{Jie Zhang}
\affiliation{Centre for Computational Systems Biology, Fudan
University, Shanghai 200433, People's Republic of China}
\affiliation{Department of Electronic and Information Engineering,
Hong Kong Polytechnic University, Hong Kong, People's Republic of
China}
\author{Xiao-Ke Xu}
\affiliation{Department of Electronic and Information Engineering,
Hong Kong Polytechnic University, Hong Kong, People's Republic of
China}
\affiliation{School of Communication and Electronic
Engineering, Qingdao Technological University, Qingdao 266520,
People's Republic of China}

\author{Michael Small}
\affiliation{School of Mathematics and Statistics, University of Western
Australia, Crawley, WA 6009, Australia}
\affiliation{Department of Electronic and Information Engineering,
Hong Kong Polytechnic University, Hong Kong, People's Republic of
China}

\date{\today}

\begin{abstract}

Recently, a framework for analyzing time series by constructing an associated complex network has attracted significant research interest. One of the advantages of the complex network method for studying time series is that  complex network theory provides a tool to describe either important nodes, or structures that exist in the networks, at different topological scale. This  can then provide distinct information for time series of different dynamical systems. In this paper, we systematically investigate the recurrence-based phase space network of order $k$ that has previously been used to specify different types of dynamics in terms of the motif ranking from a different perspective. Globally, we find that the network size scales with different scale exponents and the degree distribution follows a quasi-symmetric bell shape around the value of $2k$ with different values of degree variance from periodic to chaotic R\"{o}ssler systems. Local network properties such as the vertex degree, the clustering coefficients and betweenness centrality are found to be sensitive to the local stability of the orbits and hence contain complementary information.
\end{abstract}


\maketitle

\begin{quotation}
Methods of nonlinear time series analysis seek to quantify deterministic structure in a dynamical system from measured time series data. To date, these methods are based on dynamical systems theory and the indices of interest quantify features related to the (possible) underlying deterministic dynamics. Measures such as correlation dimension and Lyapunov exponents are widely applied and have found utility in a diverse range of sciences. The underlying principle of these methods is that the various embedding theorems provide conditions under which scalar time series measurements can be transformed into a geometric object which (under suitably generic conditions) is equivalent to the (presumed) original attractor. Recently, a new set of transformations have gained considerable interests. With these transformations it is possible to represent a scalar time series as a complex network. Previous work has already shown that these complex networks retain certain features of the original dynamical systems and therefore the arsenal of computational complex network science and theoretical graph theory may now be applied to analyse structural properties of time series data --- and by implication the underlying dynamical system. In this paper we provide the first thorough analysis of quantitative measures of the complex network properties of the recurrence-based phase space networks. We study a range of network based measures at the local-, meso- and macro-scale and show how these measures relate to the dynamical properties of the underling system. We also extend previous work by providing a rigorous computational study of the effect of noise in the time series data on the results: both for simulated and experimental data.

\end{quotation}

\section{Introduction}

Time series analysis and its applications have long been important for the
understanding of complex systems in many research fields such as sociology,
biology and society\cite{small2005applied}. Early studies mainly focused on the time series and its phase space embedding and this led to non-trivial features estimated from data: long-range correlations, scale invariance and so on. Meanwhile, a lot of recent work has  shown that natural and artificial systems in diverse fields can be also described with complex networks\cite{Albert:Statistical:2002,Barabasi:Emergence:1999,boccaletti:Complex:2006}
which are composed of a number of nodes that are interconnected by a set of
edges. Although time series and complex networks share many relevant
applications, the bridge between time series and complex network has not
appeared until Zhang and Small~\cite{Zhang:Complex:PhysRevLett.96.238701}
proposed the method of transforming pseudoperiodic time series into the
so-called cycle networks in 2006. Since then, many methods for transforming time series into complex networks have been designed, such as recurrence networks~\cite{Reik:Recurrence:2010,Donner:Ambiguities:2010,Yong:Identifying:2010,Donner:geometry:2011,Reik:Recurrence:2010}, correlation
networks~\cite{Yang:Complex:2008,Gao:Flow:2009} and visibility graphs~\cite{Lacasa:From:2008}. By going from time series to complex networks, it is possible to understand the correlation structure and dynamical properties of the time series by using the mathematical knowledge from  graph theory and complex network science.

In analyzing complex networks, one of the greatest advantages is the existence of multiple scales of identification and description of the important nodes or structures that exist in the networks.

For a given a network, even simple global statistics can provide a profound insight into the network itself. The most intensively studied ``macroscopic'' concepts are the average path length, average clustering coefficient, and degree distribution. Examples include the two well-known characterization of  scale-free and small world features widely found in  networks. The former was based on the observation that the degree distributions of many real networks have a power-law form. The latter refers to the small world phenomenon which is found especially in many social networks and is characterized by a low average path length and high average clustering coefficient.

When investigating complex networks at the ``microscopic'' scale, local properties of the network can also be important. Identifying the most vital nodes has also attracted academic research interest since these nodes could play a crucial role in the organization of the network and thus has significant impact on the dynamics of networks, such as their synchronization, epidemic spreading and so on. The proposed network statistics to measure node centrality~\cite{Albert:Internet:1999,Watts:Collective:1998,Linton:Betweenness:1977,Estrada:Subgraph:2005,Bonacich:Factoring:1972} include degree centrality, local clustering coefficient, betweenness centrality, local subgraph centrality, eigenvector centrality, closeness centrality, and so on. Though considered as local properties, these network statistics actually capture the network properties on a different scale. For example, the degree centrality which is defined as the number of neighbors of a vertex, measures only the local connections for the given node. However, the local clustering coefficient reflects the importance of the node in the network at a more macroscopic level than degree centrality since it measures the connectivity between the neighbors of a node. Similarly, the betweenness centrality of a node involves calculating the shortest paths between pairs of vertices that pass the node on a graph and therefore measures the importance of a node at an even more macroscopic level.

Moreover, much recent research attention has been put into studying the structural important features of the network, such as the motif, cyclic,
assortative patterns, and other community structures~\cite{Milo:Network:2002,Kim:Cyclic:2005,Newman:Assortative:2002,Newman:Finding:2004,Fortunato:Community:2010}. These ``mesoscopic'' structures which consist of both nodes and edges play an important structural and functional role in complex networks.

The basic idea of this paper is that the conversion from time series to complex networks provides a multiscale characterization originating from the network analysis, which in-turn gives a deeper insight into the dynamical origin of  the time series. The key questions are what information of the time series is contained inside the network representation and how may such multiple scales of identification and description be related to the time series. To address these questions, much research effort has been devoted to networks based on the important concept of recurrences from time series in the phase space. Traditionally, to construct a recurrence network following the idea of a recurrence plot proposed by Eckmann {\it et~al.}~\cite{Eckmann:Recurrence:1987} --- the time series is mapped into a set of points in the embedded phase space, each point represents a node of the network, and two nodes are linked together when their phase space distance is smaller than a selected threshold $\varepsilon$. By choosing an appropriate $\varepsilon$, the local phase space properties can be best preserved~\cite{Donner:Ambiguities:2010}. These so-called epsilon recurrence networks are phase space density dependent, and link the local properties of the network with the phase space properties of the attractor directly. This approach allows one to uncover similarities between the phase space structures of various complex systems, which can provide deeper insights into the complex systems. In 2008, Xu {\it et al.} suggested the so-called phase space networks constructed by linking each node with its $k$ nearest neighbors~\cite{Xu:Superfamily:2008} in the reconstructed phase space.
By studying the occurrence frequencies of the connected motif patterns, this methods can be used to distinguish time series with different dynamics by their so-called superfamily phenomenon.  An advantage of this approach is that the phase space networks are topologically invariant, since they depict the relative proximity relationships among a fixed number of neighboring points in the phase space attractor.

As in Xu's phase space networks (Ref.~\onlinecite{Xu:Superfamily:2008}), the network motifs which are the building blocks of the complex
networks reflect different local structure properties~\cite{Milo:Network:2002}
at the microscopic level. Besides the motifs, however, other characteristics
which are associated with local and global properties of the network and
characterize the importance of a node in the network from different scales have been left unstudied. In this paper we address this deficiency. To do so,  we will focus exclusively on the phase space network of Ref.~\onlinecite{Xu:Superfamily:2008} and not the recurrence methods of Refs.~\onlinecite{Donner:Ambiguities:2010},~\onlinecite{Reik:Recurrence:2010} and~\onlinecite{Marwan:Complex:2009}.
For an extensive comparison of these methods, and the other emerging alternatives, see
Ref.~\onlinecite{Donner:Recurrence:2011}.

\begin{figure}[!ht]
 \centering
\includegraphics[trim=0 0 0 0, clip=true, scale=0.75]{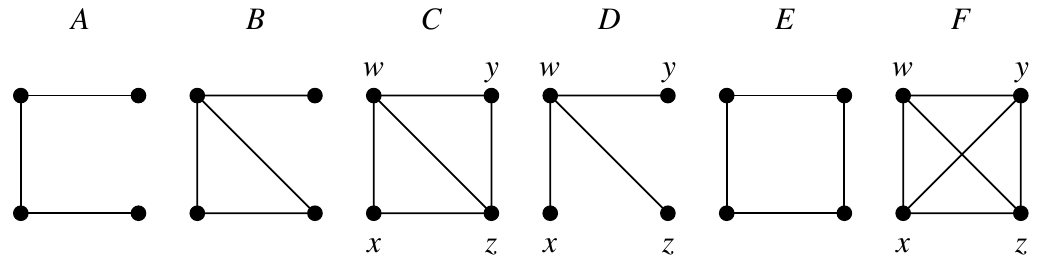}
\parbox{8cm}{\caption{\label{fig:motif}
All six motifs of size four denoted by $A$, $B$, $C$, $D$, $E$ and $F$.}}
\end{figure}

The reminder of the paper is organised as follows: we first quantify these recurrence-based phase space networks (using the adaptive nearest neighbor network method~\cite{Xu:Superfamily:2008,Donner:Recurrence:2011}) via the standard topological statistics, in addition to the motif ranking, to show how such global and local network statistics can provide new information about the phase space geometry beyond time series. Then we apply the same analysis to noisy experimental data from continuous musical tones produced on a standard clarinet. In the last section of this paper we provide a brief conclusion.

\section{\label{sec2:}Quantitative assessment of recurrence-based phase space networks}


The recurrence-based phase space network representative for a given time series is generated following the same method used in Ref.~\onlinecite{Xu:Superfamily:2008} and summarized in Ref.~\onlinecite{Donner:Recurrence:2011}. The first step is to embed the time series in an appropriate phase space. After proper embedding, each point in the phase space is taken as a node of the network and will be assigned with $k$ new nearest neighbors. Eligible $k$ neighbors should exclude points within the same orbit as well as those which are already neighbors. On the one hand this ensures that the neighbors are spatial neighbors in the phase space rather than the temporal neighbors, on the other hand this avoids multiple links between two nodes and provides a measure of phase space density. By constructing network in this way, we are able to enforce a
threshold adaptively according to the local inter-point density of the attractor and thus obtain a network representative of time series which inherits structure of the distinct local phase space properties from different dynamical systems.


For a low-period periodic time series, the associated network is regular as the points are arranged in an orderly manner around an orbit in the phase space. The network forms a large loop shape and its size increases proportionately to the number of points of the embedded data. If we add noise to the data, the dimension of the dynamics increases while the homogenous distribution of the points remain unchanged: we still get the loop structure, but the diameter of the network decreases. For chaotic time series, the reconstructed phase space becomes heterogenous and may have some fractal properties and thus thus the structure of the associated network is more heterogeneous.

\begin{figure}[!ht]
 \centering
\includegraphics[trim=0 0 0 0, clip=true, width=0.45\textwidth]{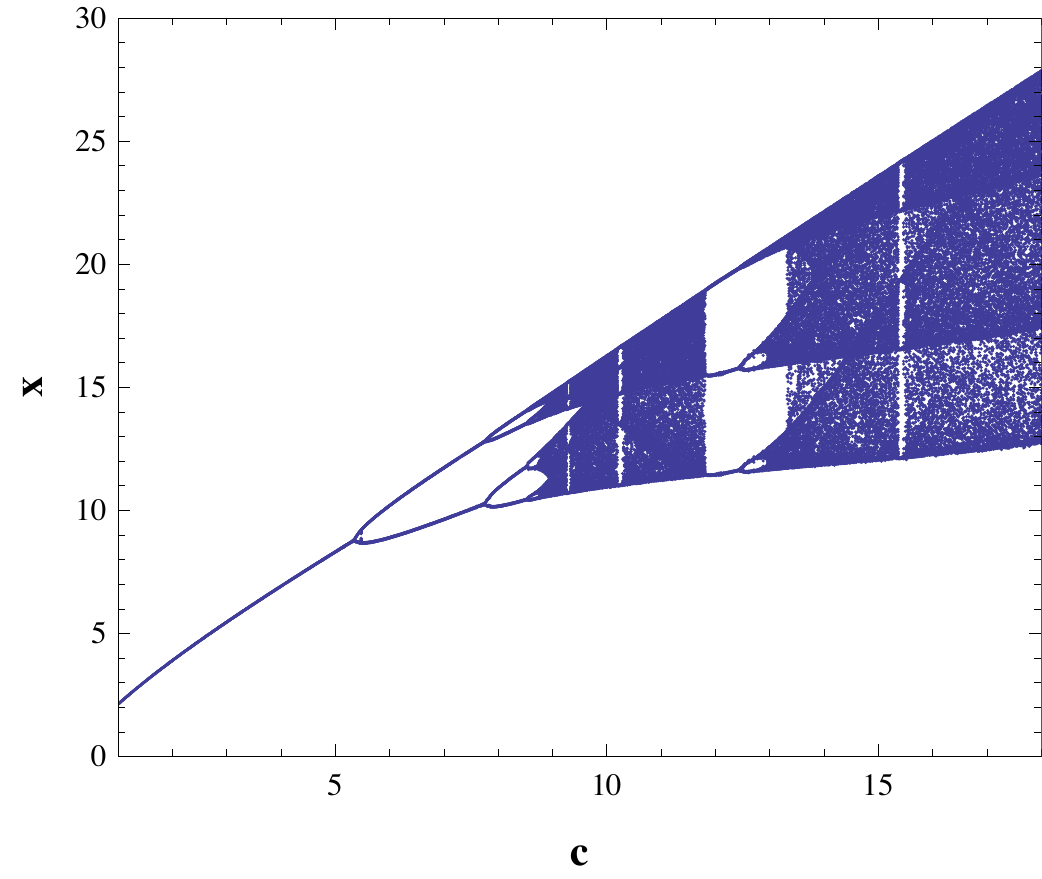}
\parbox{8cm}{\caption{\label{fig:bifurcation}
Bifurcation diagram illustrating the dependence of $x$ component of R\"{o}ssler system on the parameter $c$.
}}
 \end{figure}

As a typical example for a continuous dynamical system, we study the R\"{o}ssler system. We make use of the data from the x component of the R\"{o}ssler system by solving $x'=-(y+z)$, $y'=x+ay$, $z'=b+(x-c)z$ with $a=0.1$ and $b=0.1$, using the fourth--fifth order Runge--Kutta method (ODE45 in MATLAB) with a fixed sampling rate of $h=0.1$. By tuning the parameter $c$, low-period periodic, high-period periodic and chaotic data can be obtained.
Figure~\ref{fig:bifurcation} shows the bifurcation diagram, with parameter $c\in[1,18]$ divided into $1000$ intervals.
The data is mapped into a space of dimension $d_e=10$ that is sufficiently large to study the topological structure with the time delay $\tau$ which is chosen to be the first minimum of the mutual information. To  construct a network from the time series, we take each point in the embedded space as a node and link it with its $4$ nearest neighbors which will lead to a network with the mean degree equal to 8. The choice of $4$ neighbors follows Ref.~\onlinecite{Xu:Superfamily:2008} to guarantee a robust results --- however, as noted in that reference, this choice does not significantly affect the results. Nonetheless, the number of nodes of the network is equal to $N-\tau\cdot (d_e-1)$ where $N$ is the time series length.

 \begin{figure}[!ht]
 \centering
\includegraphics[trim=0 0 0 0, clip=true, width=0.45\textwidth]{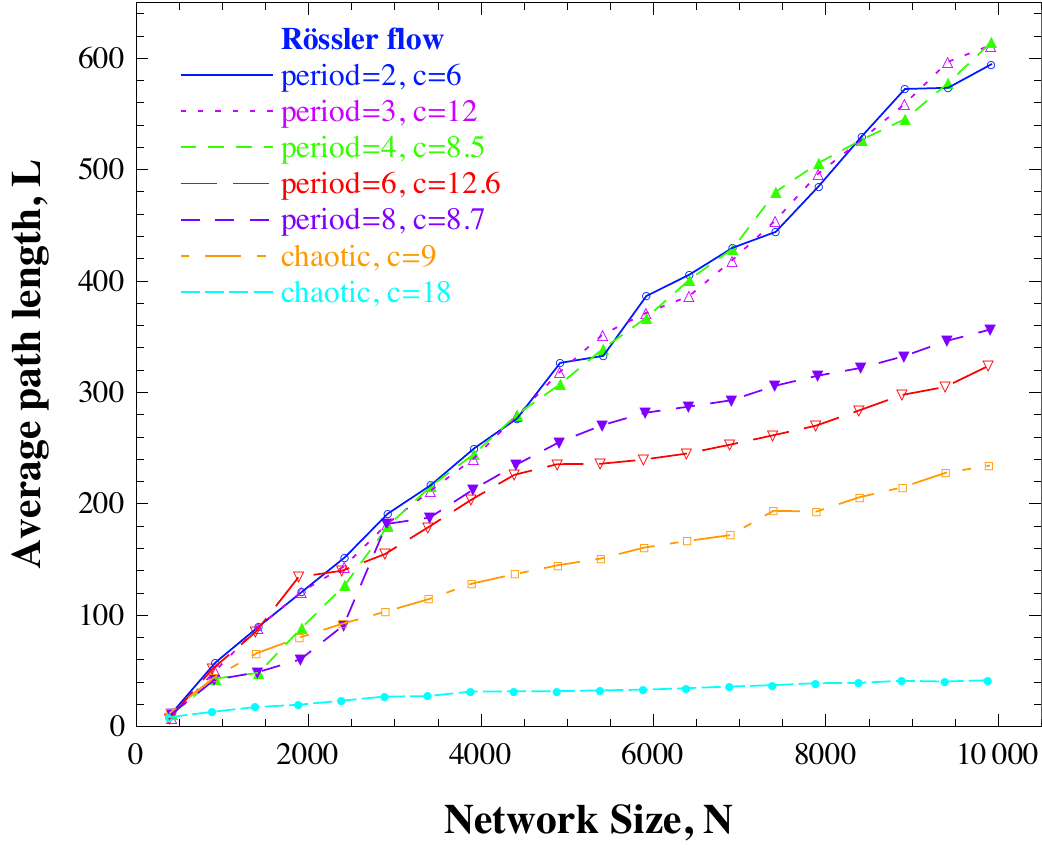}\\
\parbox{8cm}{\caption{\label{fig:APL}
    Dependence of the average path length $L(N)$ on the length $N$ of
the time series for the R\"{o}ssler system in different dynamical regimes.
 Best fitting models are $L(N)=0.085 N^{0.96}$ and $L(N)=0.059
N^{1.00}$ for the low-order periodic R\"{o}ssler system of period 2 and 3
respectively. In comparison, best fitting models are $L(N)=0.48 N^{0.67}$ and
$L(N)=0.78 N^{0.43}$ for the chaotic R\"{o}ssler system with $c=9$ and $c=18$
respectively. }}
 \end{figure}

\subsection{Global network characteristics}
\subsubsection{Average path length}
To investigate the distinction between qualitatively different types of dynamics in more detail, we first study the average path length of the phase space network as a function of the time series length. Figure~\ref{fig:APL} shows how the average path length of the phase space network changes with respect to the increasing length of the time series. Note that for shorter time series, the average path length can be affected by the choice of the initial point of the flow data, so the desired results are taken as an average over several networks from different segments of the same parameter.  We find that the average path length $L(N)$ scales with $N$ with different scaling exponents for low-order periodic and chaotic R\"{o}ssler systems. The average path length $L(N)$ increases linearly with time series lengths $N$ for low-order periodic time series with periods 2 and 3, but exponentially for the chaotic R\"{o}ssler system. For high-order periodic cases, we find transient behavior when the size $N$ is small because the corresponding time series can be too short to cover enough information with respect to the periodicity. Moreover, the slopes for the average path length $L(N)$ are larger than the two chaotic R\"{o}ssler cases but smaller than the low-order periodic cases due to the interactions between two nearby orbits appearing in the phase space.

 \begin{figure}[!ht]
 \centering
 \includegraphics[trim=0 0 0 0, clip=true, width=0.45\textwidth ]{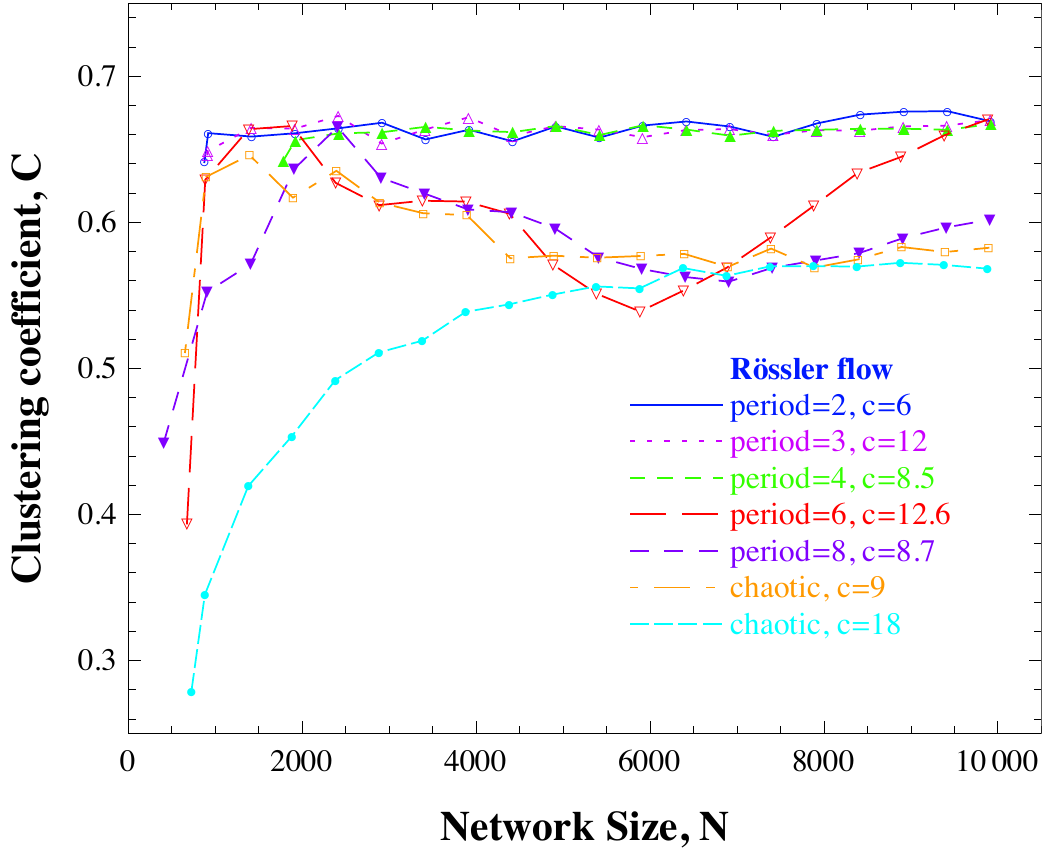}
 \parbox{8cm}{\caption{\label{fig:cc}
     Dependence of the global clustering coefficient $C(N)$ on the length N of
the time series.}}
 \end{figure}

We also look into other global properties including the clustering coefficient and the degree distribution for R\"{o}ssler systems of different dynamical type (chaotic, noisy or periodic motion).
\subsubsection{Clustering coefficient}
The global clustering is closely related to motif ranking. Motif $F$, which denotes a fully-connected subgraph of order 4 as shown in Fig.~\ref{fig:motif}, indicates strong mutual coupling between nodes, and hence we may expect that the clustering coefficient for the periodic R\"{o}ssler will be larger than the chaotic R\"{o}ssler, because motif $F$ occurs more frequently in periodic phase space networks rather than networks from chaotic data of the same R\"{o}ssler system as mentioned in Ref.~\onlinecite{Xu:Superfamily:2008}. From Fig.~\ref{fig:cc} we can see that the clustering coefficients are rather stable with respect to an increasing $N$ and converges to around 0.68 for the low-order periodic regimes with periods of $2$ or $3$ as the network size increases. Similar behavior can be found in the chaotic R\"{o}ssler system with $c=18$. The clustering coefficients are also stable but converge to a lower value around 0.58. In comparison, significant transient effect can be found in the remaining three cases. The curve for the chaotic R\"{o}ssler system with $c=9$ first rises to the same high level as in the low-order periodic cases for the $N$ between 1000 and 2000. Then, it decreases and finally converges to a low level similar to the chaotic case. The curve for the R\"{o}ssler system of period 6 goes up and down and finally approaches the same value as the low-order periodic cases. Although the value of the clustering coefficient for the R\"{o}ssler system of period 8 is closer to that of the chaotic cases, we may infer that, as the size of the
network increases, the curve will finally converge to the low-order periodic values.

 \begin{figure}[!ht]
 \includegraphics[trim=10 0 0 0, clip=true, width=0.45\textwidth]{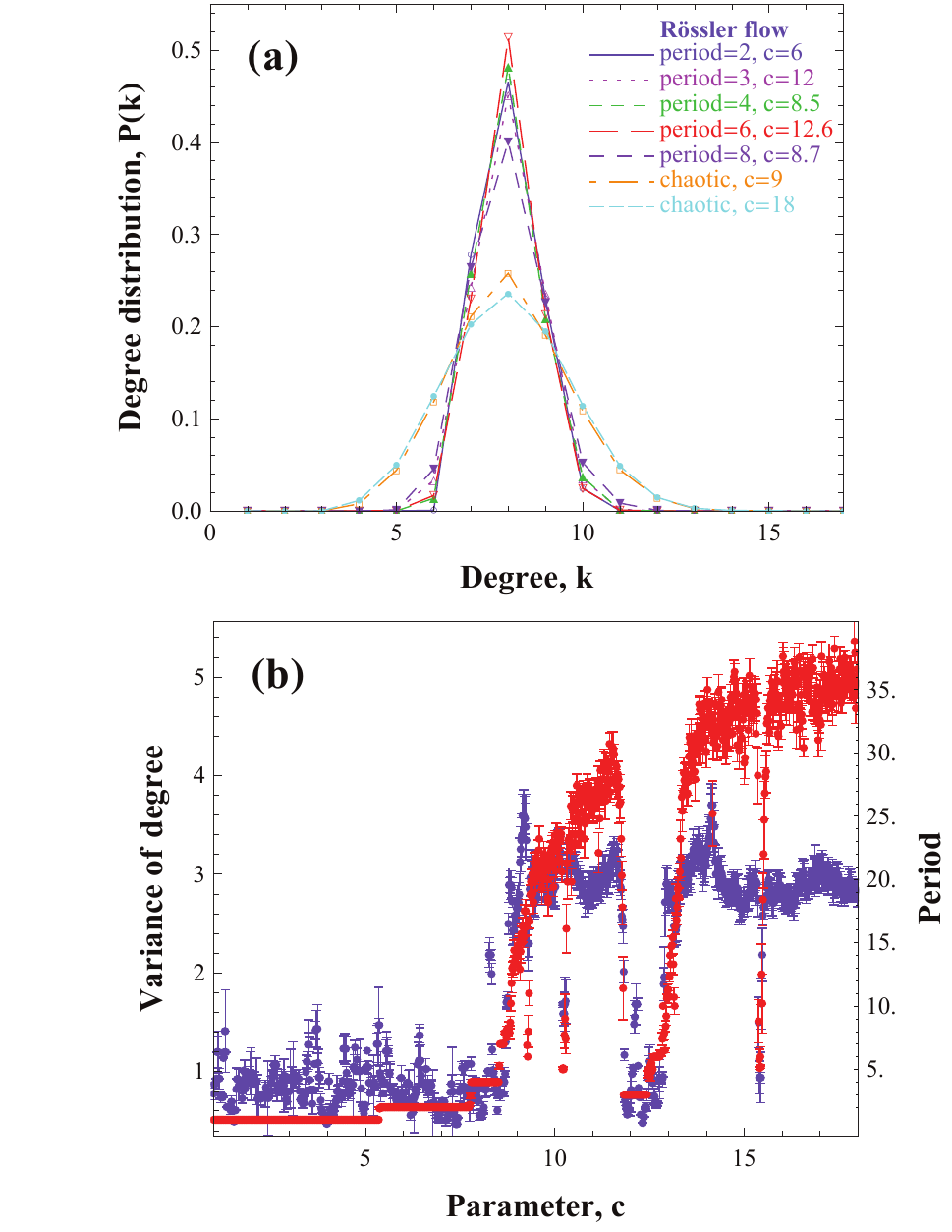}\\
 \parbox{8cm}{\caption{\label{fig:dg}
    (a) The degree distribution for the R\"{o}ssler system of different dynamical types. (b) Red line: The calculated periods according to different values of parameter $c$. Blue line: The variance of degrees with respect to the change of parameter $c$.  }}
 \end{figure}

\subsubsection{Degree distribution}
Degree distribution itself measures the global network connectivity and  can therefore capture the global network topological properties inherited from the time series. Figure~\ref{fig:dg}(a) shows the average degree distributions for networks of several realizations for different initial conditions associated with 10000 points of the R\"{o}ssler system with different dynamics. We find that the average degree distribution follows a quasi-symmetric bell-shape for networks from the R\"{o}ssler system in different dynamical regimes. The degree at which $p(k)$ reaches its maximum is 8, which is exactly the average degree of the network (i.e. mode and median are equal). The peak values of $p(k)$ for networks generated from periodic data are all above 0.3 and the variance of the degrees with bell-shaped distribution are rather small, indicating that the networks are homogenous. In contrast, for the chaotic R\"{o}ssler system, $p(k)$ is lower than 0.3 and the variance of the degrees are larger.

We plot the variance of degrees in the phase space network as a function of the bifurcation parameter $c$ in Fig.~\ref{fig:dg}(b) and find a much clearer relationship between the bifurcation diagram and the variance of degrees in the network. We make use of 10 realizations from random initial points. For each of the 10 realizations, there are 10000 points for every step of $c$. In Fig.~\ref{fig:dg}(b) the red line shows the period while the blue line displays the variance of degrees with respect to the gradual changes of parameter $c$. To get the periods as a function of the parameter $c$, we first detect the locations of maximum points at cycle scale. If the difference of any two maximum points is larger than a given threshold, we group them into different periodic orbits. Finally the number of orbits is considered as the period of the data. Due to the limit of numerical precision, if the numerical number of periods exceeds a certain large value, then that is taken to be an indicator of chaos (that is, we are forced to consider aperiodicity and periodicity with period longer than some arbitrary large threshold as equivalent). This simple numerical method reveals the bifurcation scenario of the period-doubling route to chaos with the occurrence of several periodic windows in between, depicted by the red line. For comparison, we also show the average variance of degrees of the corresponding networks. It is clear that this variance is in general larger in the chaotic region. Note that when an incommensurate sampling frequency is chosen, the associated recurrence-based phase space network could become disconnected and be divided into several components rather than a regular single network at some values of parameter $c$ in periodic region, inevitably leading to large variance in degree which appears in even low-order periodic regions of the bifurcation diagram. Hence, we use a moving average with window width 3 for
consecutive values to smooth the curve of the calculated values for variance of degrees in the blue line in Fig.~\ref{fig:dg}(b).


 \begin{figure}[!ht]
 \centering
 \includegraphics[trim=0 0 0 0, clip=true, scale=0.75]{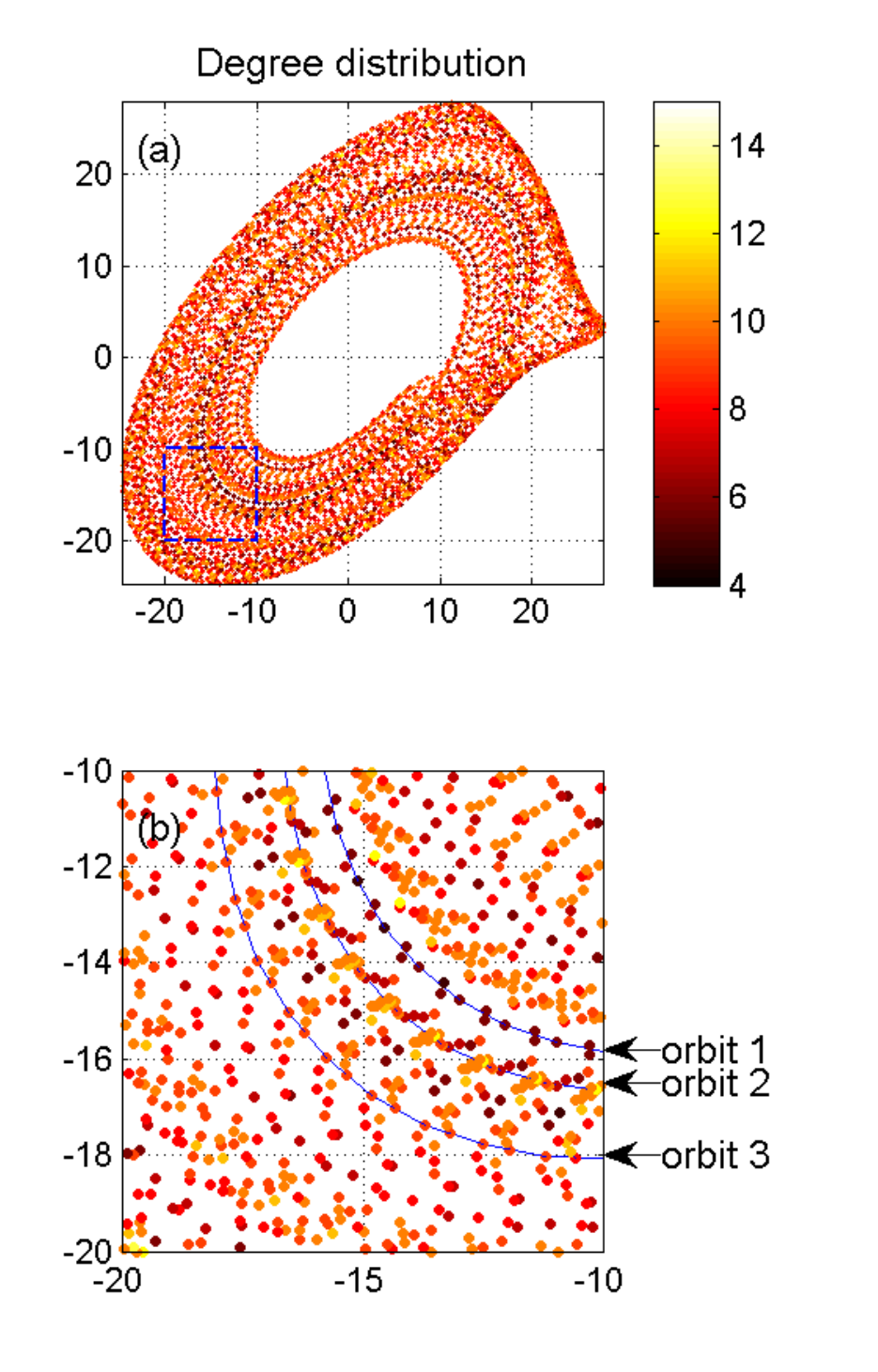}\\
 \parbox{8cm}{\caption{\label{fig:dglc}
   (a) Color coded representation of vertex degree on the attractor for the chaotic
R\"{o}ssler system with $c=18$. (b) Zoom-in area in the dashed box of (a).}}
 \end{figure}

\subsection{Local network characteristics}
\subsubsection{Local vertex degree}
Rather than simply looking at the global properties of the network, we also examine local vertex properties which we believe can provide more detailed information that is otherwise buried in the average geometry of the attractor. The first local property we study is the local vertex degree which is the number of neighbors of a given vertex $v$ in the network. The degree of a vertex for the traditional recurrence networks which are based on an appropriate choice $\varepsilon$ reflects exactly the local recurrence rate. When it comes to the phase space network, we should recall that the average degrees are 8 due to the specific construction methods we use to obtain the network. Thus the local degrees depend closely on density gradients of the phase space attractor. On the one hand, in the region where the
density of the phase space is homogenous the corresponding structures of the network retain the homogeneity from the attractor and local degrees for the vertices will be around 8. That is, every point in these regions is connected to its 4 nearest neighbors and at the same time gains (on average) 4 extra links from points which take it as a nearest neighbor. One the other hand, if points are located in the regions of large gradient in the phase space density, some of these points may gain more than 4 extra links besides their 4 nominal nearest neighbors, resulting in the lack of neighbors for some other points in the same region. These regions will have a heterogeneous degree
distribution. Long periodic signals follow a uniform distribution of low
dimension while the trajectories of chaotic signals tend to be trapped in the
vicinity of the unstable periodic orbits (UPO) that lead to large gradients in the phase space density. One may refer to color-coded representation of the vertex degree on the attractor for the chaotic R\"{o}ssler system with $c=18$ in Figs.~\ref{fig:dglc}(a) and (b) for details. Overall, we get the heterogeneously  distributed degrees in the attractor in Fig.~\ref{fig:dglc}(a). Typically, there are UPOs in the chaotic attractor. The trajectory will approach a UPO along its stable manifold, stay close to the UPO for several cycles and then be ejected away until it is captured by the stable manifold of another UPO, leading to density variation.

\begin{figure}[!ht]
 \centering
 \includegraphics[trim=0 0 0 0, clip=true,width=0.45\textwidth]{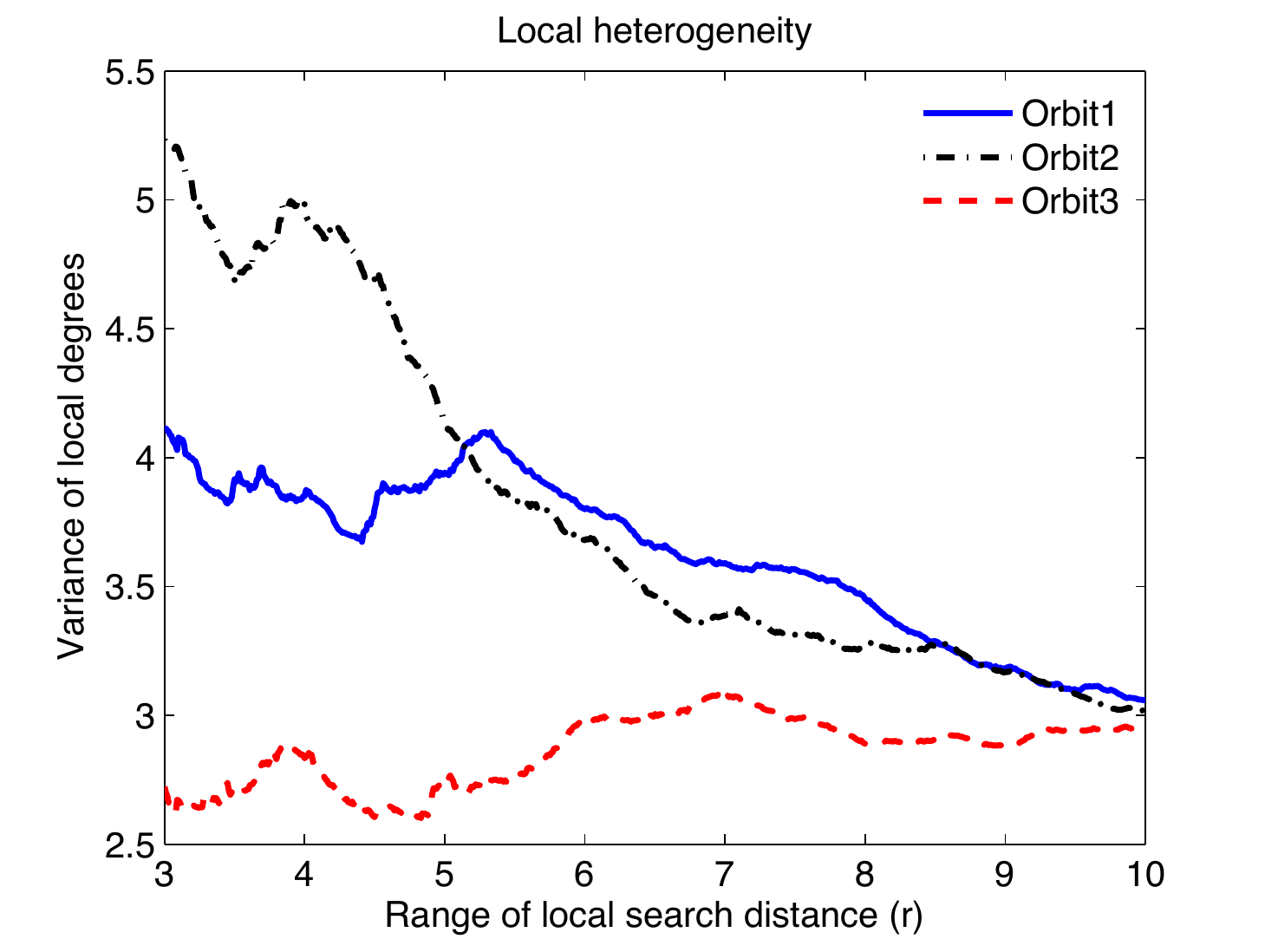}\\
 \parbox{8cm}{\caption{\label{fig:localheterogeneity}
   Local heterogeneity in terms of degree variance for the three orbits.}}
 \end{figure}

To provide a more quantitative analysis, we introduce the local heterogeneity for a specific part of phase space which is defined by the variance of degrees of all points located within a ball of appropriate radius $r$ in the phase space. By zooming in a particular region of the attractor showed in Fig.~\ref{fig:dglc}(b), we check three orbits representing trajectories in different situations with local heterogeneity plotted in Fig.~\ref{fig:localheterogeneity}. Orbit 1 is in a low-density region while orbit 2 is located in the region close to a dominant UPO with a high density in this chaotic attractor. We observe a heterogeneous degree distribution in the nearby trajectory along orbits 1 and 2, because both of the two orbits are located in regions with higher gradient of attractor density. In comparison, we get a homogenous degree distribution in the region along orbit 3, which is located in the part of the attractor with lower gradient in density.  If the radius $r$ is chosen too large, the ball will cover a larger part of the attractor and local heterogeneity will be removed. Hence we find that the three curves in Fig.~\ref{fig:localheterogeneity} tend to converge to a fixed value as $r$ increases.
Moreover, we can address the diverse variance of degrees in Fig.~\ref{fig:dg} for different dynamical types. Since the attractor becomes more heterogeneous  as the dynamics changes from low-order periodic to high-order periodic and finally chaotic, the values for the degree variance will be larger.


 \begin{figure}[!ht]
 \centering
 \includegraphics[trim=0 0 0 0, clip=true, scale=0.75]{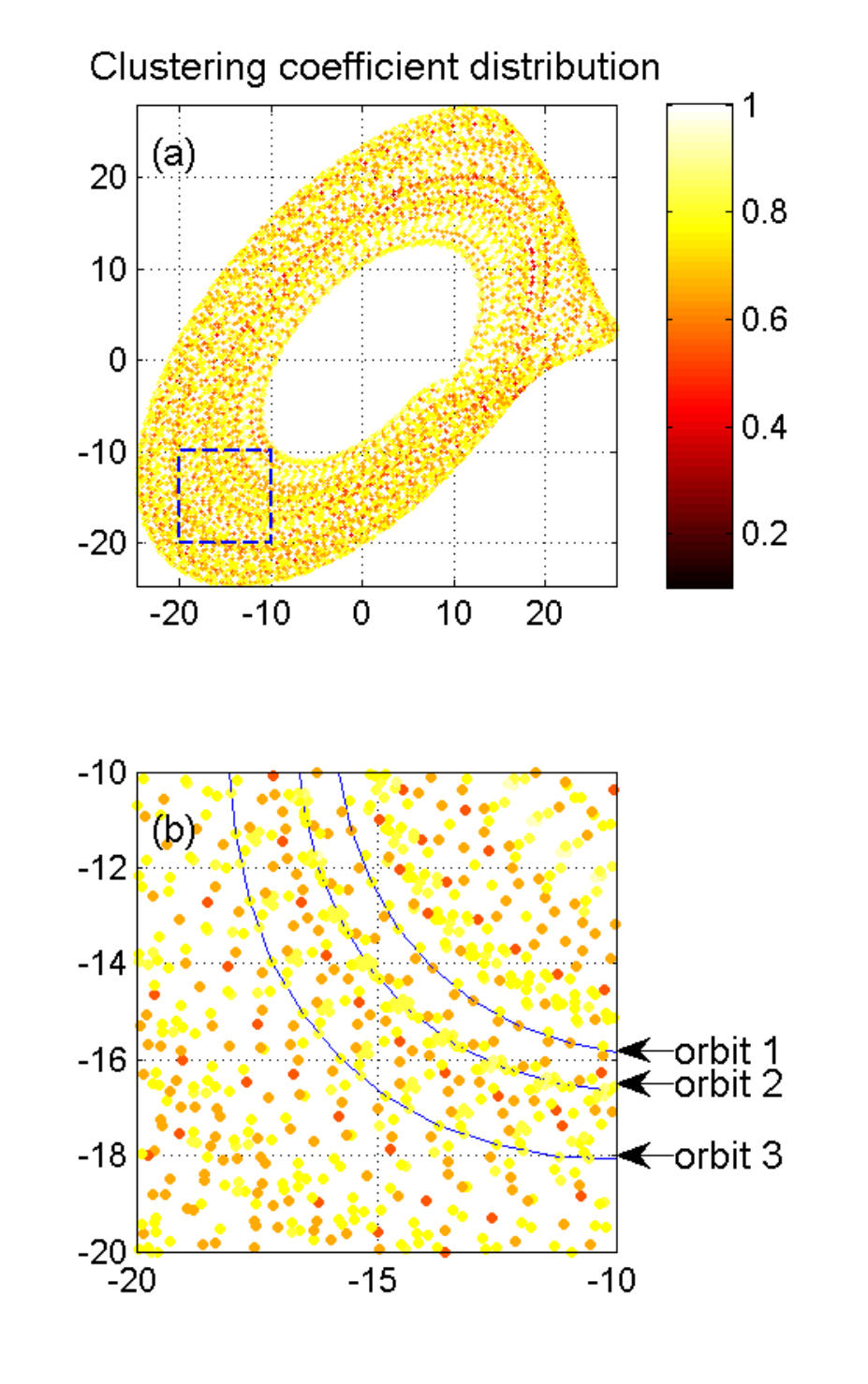}
 \parbox{8cm}{\caption{\label{fig:cclc}
    (a) Color coded representation of the vertex clustering coefficient on the
attractor for the chaotic R\"{o}ssler system with $c=18$.(b) Zoom-in area in the dashed box of (a).}}
 \end{figure}

\subsubsection{Clustering coefficient of a specific vertex}
The second property we study is the clustering coefficient of a specific vertex which quantifies the relative density of connections between its neighbors and
thus measures the network at a relatively large scale compared to the local
degree. Figures~\ref{fig:cclc}(a) and (b) show color representations of the vertex clustering coefficient on the attractor for the chaotic R\"{o}ssler system with $c=18$, from which we observe the heterogeneously distributed local clustering coefficients. One might expect a similar effect for epsilon-recurrence networks within a certain range where a node in a dense region of the phase space will naturally have a larger clustering coefficient and vice versa. However, for recurrence-based phase space network, shown here in Figs.~\ref{fig:cclc}(a) and (b), the values of the clustering coefficient for some nodes located in the denser regions of the attractor may be small. For example, clustering coefficients for the nodes located on the nearby orbit next to orbit 2 are small, while the values for nodes located in the boundary region of the attractor can be large. In a homogeneously filled region, things become simpler and we can see points with relatively uniform and higher clustering coefficients. However, when it comes to a region with large density fluctuations, we have to consider the interactions between several orbits which are close to each other. A strong connectivity between neighbors of the same point can be obtained only if its neighbors tend to lie in different orbits, because we have excluded links within the same orbit when constructing the phase space networks\footnote{That is, we explicitly exclude the possibility that two nodes may be connected if the corresponding time series points have a temporal separation less than half the system pseudo-period.}. We can conclude that the local clustering coefficient contains the complexity of the interaction between orbits in the phase space and can provide complementary information on the system.

 \begin{figure}[!ht]
 \centering
 \includegraphics[trim=0 0 0 0, clip=true, scale=0.75]{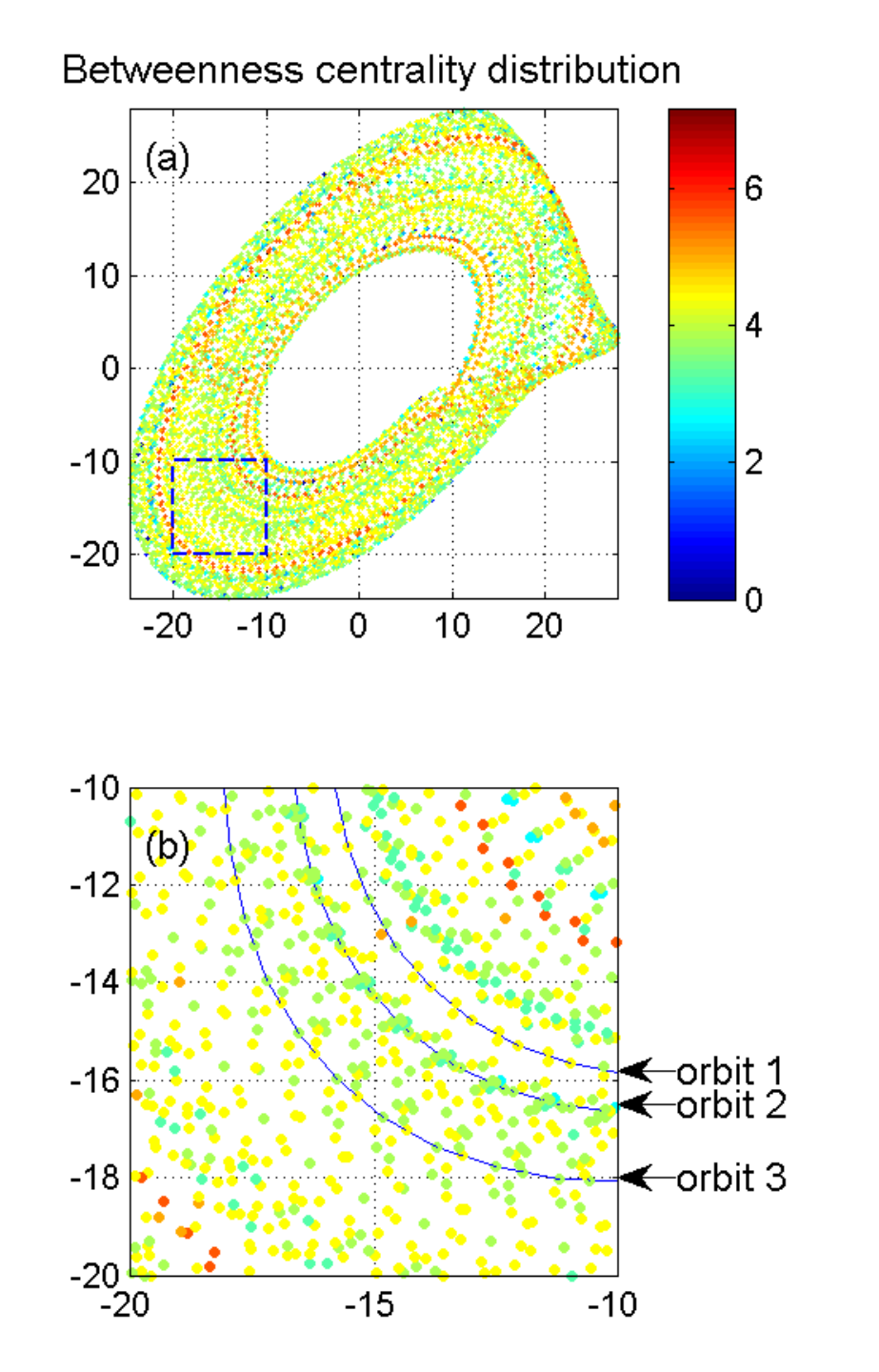}
 \parbox{8cm}{\caption{\label{fig:bclc}
   (a) Color coded representation of the logarithm of vertex betweenness centrality, $\log(b_v+1)$, on the attractor for the chaotic R\"{o}ssler system with $c=18$. (b) Zoom-in area in the dashed box of (a).}}
 \end{figure}

\subsubsection{Betweenness centrality of a specific vertex}
The third property we study is the betweenness centrality of a specific vertex
which quantifies the frequency with which shortest paths (between a randomly chosen pair of nodes) pass through the vertex. Thus it is more complicated because it measures the network at an even larger scale compared to the clustering coefficient and the local degree. What we are most interested in are the points with a large betweenness centrality which are of importance for many shortest paths on the network. As shown in Fig.~\ref{fig:bclc}, the nodes with high betweenness centrality may correspond to a region with low phase space density between such regions because of an increasing number of shortest path between such regions.

 \begin{figure}[!ht]
 \centering
 \includegraphics[trim=0 0 0 0, clip=true,width=0.45\textwidth]{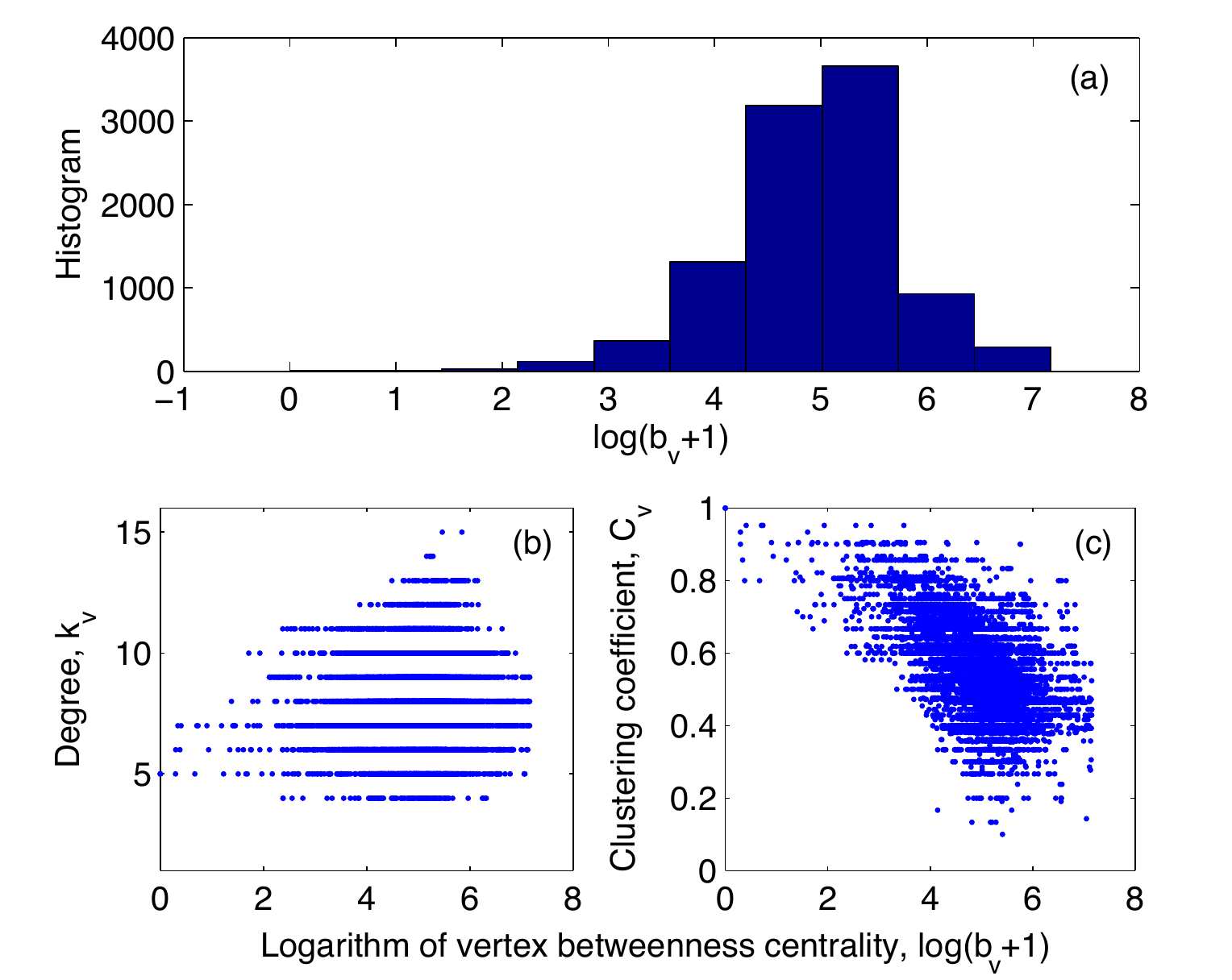}
 \parbox{8cm}{\caption{\label{fig:bc}
   (a) Histogram for the logarithm of vertex betweenness centrality, $\log(b_v+1)$. (b) Relationships between local degree, $k$ and logarithm of betweenness centrality, $\log(b_v+1)$.  (c) Relationships between local clustering coefficient, $C_v$ and logarithm of betweenness centrality, $\log(b_v+1)$.}}
 \end{figure}

In Fig.~\ref{fig:bc}, we show the relationship between various local network properties. In Fig.~\ref{fig:bc}(a), the log-betweenness follows a bell-shaped distribution in which most of the points have around $10^4-10^5$ shortest paths passing through. From Figs.~\ref{fig:bc}(a) and (b) which shows the relationships of vertex betweenness centrality with local degree and local clustering coefficient, points with low betweenness (below $10^2$) have relative low degree and high clustering coefficient. In both relationships, however, we find a broad range of values of betweenness for most of points with respect to a fixed value of either degree or clustering coefficient.


As a consequence of the above observations, the complex
network scheme can be used as a multi-scale tool to
describe properties contained in time series from global
to local scales. Moreover, different local vertex properties
contain information on different aspects from the geometry of the attractor.

\section{\label{sec3:noise}Effect of noise}

We now extend the multiscale characterization of complex networks obtained from time series to systems corrupted by noise. The first question here is: how will the noise affect the super-family phenomenon shown in Ref.~\onlinecite{Xu:Superfamily:2008}? To look for the answer, we systematically calculate the motif ranking for data from a different dynamical system corrupted by different levels of white Gaussian noise. The results are shown in Table~\ref{tbl:motifnoise}. When checking the first column in Table~\ref{tbl:motifnoise} which shows results for signals without noise, we obtain the superfamily phenomenon which is consistent with the results shown in Ref.~\onlinecite{Xu:Superfamily:2008}. We keep the embedded parameters the same as for signals which are free of noise, because we want to investigate the effect of noise on the network structure rather than that of the embedding parameters. Networks from both continuous dynamics and map data exhibit superfamily phenomena that we can identify as periodic, chaotic, hyperchaotic flow and maps by using the phase space method. For example, motif ranks for networks from the periodic R\"{o}ssler system are $ABCFDE$. Some networks completely lack motif $D$ or $E$ because the frequencies of the two are statistically insignificant for those networks. Chaotic flows such as Chua, Lorenz and chaotic R\"{o}ssler systems with $c=18$ exhibits the same motif rankings $ABCDFE$. The motif ranking for Mackey-Glass flow is $ADBCEF$ which is similar to periodic systems corrupted by noise because the system is in a high-dimensional state. For map data, the three chaotic maps (Henon map, Ikeda map and logistic map) all have motif ranks $ABCFDE$, while the hyperchaotic generalized Henon map and folded-tower map data have the same subgraph ranks $ABCDFE$ and hence belong to one superfamily.

We find that the superfamilies become less distinct when we gradually increase the amplitude of noise --- once the noise becomes too strong, motif superfamilies lose their discriminatory power, as expected.  By applying an intermediate amount of noise we can see the switch in the sequence of motifs which shows the transitions from pure signals with no noise to signals with levels of noise. The rank of nontransitive motifs of type $D$ representing a subgraph with 4 nodes and 3 edges which connect one node to the other three nodes is increasing, indicating that the system is higher-dimensional, or the distribution of nodes is more heterogeneous. By checking the last column in Table~\ref{tbl:motifnoise} we see results for signals with very significant noise ($0$dB noise). Here we observe that the networks are divided into two groups by motif ranks: all the flow data have motif ranks $ADBCEF$, while all the map data have motif ranks $ABDCEF$. Of course, for more moderate levels of noise, the power of the motif superfamily classification to correctly distinguish different dynamics improves.  Finally, we observe the possibly surprising result that adding a higher level of noise actually helps to distinguish between maps and flows. Although this distinction is artificial (all the data we consider are by definition discretely sampled), it is effectively a measure of the relative sampling rate of the dynamics and the discrete noise source. Systems with smooth dynamics and discontinuous dynamical systems become easy to distinguish when details of dynamical types are obscured by noise.

\begin{table*}[htbp]
\caption{\label{tbl:motifnoise}
Motif ranking for system corrupted by different levels of noise. }
\begin{center}
\begin{tabular}[c]{|l|l||l|l|c|c|c|}\hline
  System &Dynamical type& No noise &30dB   & 20dB
  & 10dB  & 0dB \\[0.5ex]\hline\hline
R\"{o}ssler $c=6$	 	&period$=2$		    &$ABCFD$	&$ABDCEF$	&$ADBCEF$	 &$ADBCEF$	&$ADBCEF$\\\hline
R\"{o}ssler $c=12$	    &period$=3$  		&$ABCF$&	$ABDCFE$	&$ABDCEF$	 &$ADBCEF$	&$ADBCEF$\\\hline
R\"{o}ssler $c=8.5$	    &period$=4$	    	&$ABCFD$	&$ABDCFE$	&$ADBCEF$	 &$ADBCEF$	&$ADBCEF$\\\hline
R\"{o}ssler $c=12.6$	&period$=6$		    &$ABCDFE$	&$ABDCFE$	&$ABDCEF$	 &$ADBCEF$	&$ADBCEF$\\\hline
R\"{o}ssler $c=8.7$	    &period$=8$		    &$ABCDFE$	&$ABDCFE$	&$ADBCEF$	 &$ADBCEF$	&$ADBCEF$\\\hline
R\"{o}ssler $c=9$		&chaotic flow    	&$ABCFDE$	&$ABDCFE$	&$ADBCEF$	 &$ADBCEF$	&$ADBCEF$\\\hline
R\"{o}ssler $c=18$	    &chaotic flow       &$ABCDFE$	&$ABCDFE$	&$ABDCEF$	 &$ADBCEF$	&$ADBCEF$\\\hline
Lorenz			        &chaotic flow      	&$ABCDFE$	&$ABCDFE$	&$ABDCFE$	 &$ADBCEF$	&$ADBCEF$\\\hline
Chua			     	&chaotic flow      	&$ABCDFE$	&$ABDCFE$	&$ADBCEF$	 &$ADBCEF$	&$ADBCEF$\\\hline
Mackey-Glass	       	&chaotic flow      	&$ADBCEF$	&$ADBCEF$	&$ADBCEF$	 &$ADBCEF$	&$ADBCEF$\\\hline
Logistic			    &chaotic map     	&$ABCFDE$	&$ABDCFE$	&$ABDCEF$	 &$ABDCEF$	&$ABDCEF$\\\hline
Henon				    &chaotic map    	&$ABCFDE$	&$ABDCFE$	&$ABDCEF$	 &$ABDCEF$	&$ABDCEF$\\\hline
Ikeda		     		&chaotic map       	&$ABCFDE$	&$ABDCEF$	&$ABDCEF$	 &$ABDCEF$	&$ABDCEF$\\\hline
hyperHenon              &hyperchaotic map	&$ABCDFE$	&$ABCDFE$	&$ABDCFE$	 &$ABDCEF$	&$ABDCEF$\\\hline
hyperTower              &hyperchaotic map	&$ABCDFE$	&$ABDCFE$	&$ABDCFE$	 &$ABDCEF$	&$ABDCEF$\\\hline
\end{tabular}
\end{center}
\end{table*}

\begin{figure}[!ht]
 \centering
\includegraphics[trim=0 0 0 0, clip=true, width=0.45\textwidth]{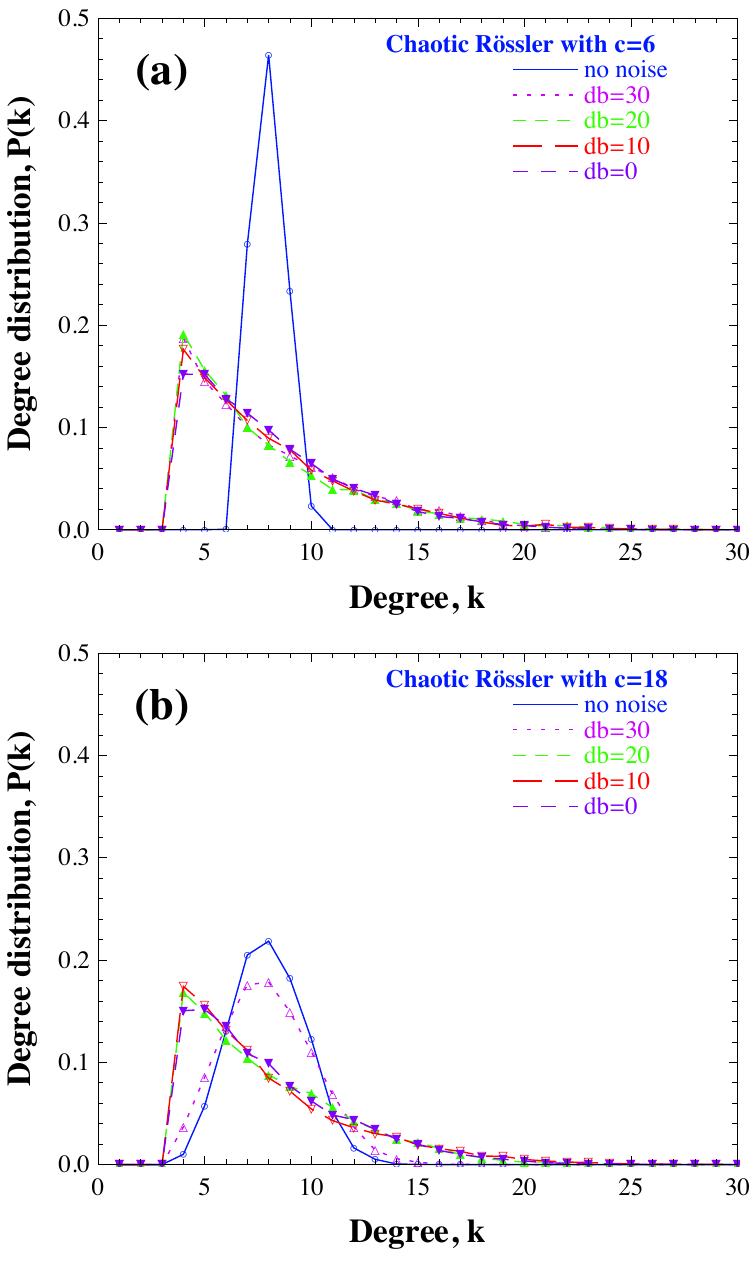}
\parbox{8cm}{\caption{\label{fig:dgnoise}
Degree distribution for the chaotic R\"{o}ssler system with additive noise of different levels from no noise to $0$dB white Gaussian noise with (a) $c=6$ and (b) $c=18$.
}}
\end{figure}

We also examine the impact of noise on the standard measurements of the network including degree distribution, average path length and clustering coefficient. For the global degree distribution that we have discussed above, we infer from the network construction methods that the average degree remains 8. In the presence of noise as shown in Fig.~\ref{fig:dgnoise}, however, the bell shape degree distribution loses its symmetry and becomes a skew distribution with mode 4.

\begin{figure}[!ht]
 \centering
\includegraphics[trim=0 0 0 0, clip=true, width=0.45\textwidth]{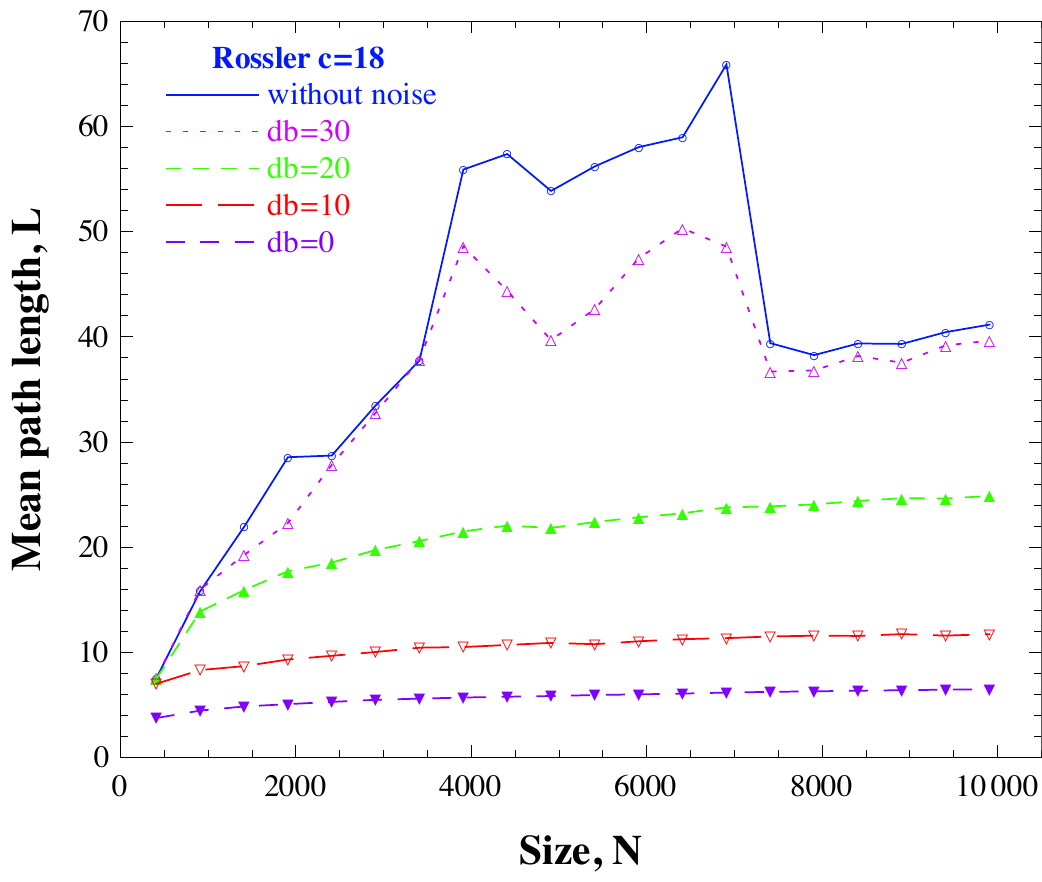}
\parbox{8cm}{\caption{\label{fig:APLnoise}
Average path length for the chaotic R\"{o}ssler system with $c=18$ contaminated by different noise levels.
}}
\end{figure}

In Fig.~\ref{fig:APLnoise}, we show how the average path length changes the network size for a particular realization in the presence of noise of different levels. The scaling behavior for the average path length has been shown in the previous section. In comparison, for a particular run, only when the length of the signal is long enough can it contain sufficient phase space information, for example enough UPOs, to characterize chaos. Hence the average path length is larger than the expected scaling value for networks from signals with length smaller than 7000. This does not affect the result, however, as we only try to explore the impact of noise. In the presence of a small amount of noise, the average path length decreases a little. When we increase the noise level, we still see the scaling curve increasing, but the average paths length becomes much smaller than the original one.

\begin{figure}[!ht]
 \centering
\includegraphics[trim=0 0 0 0, clip=true, scale=0.7]{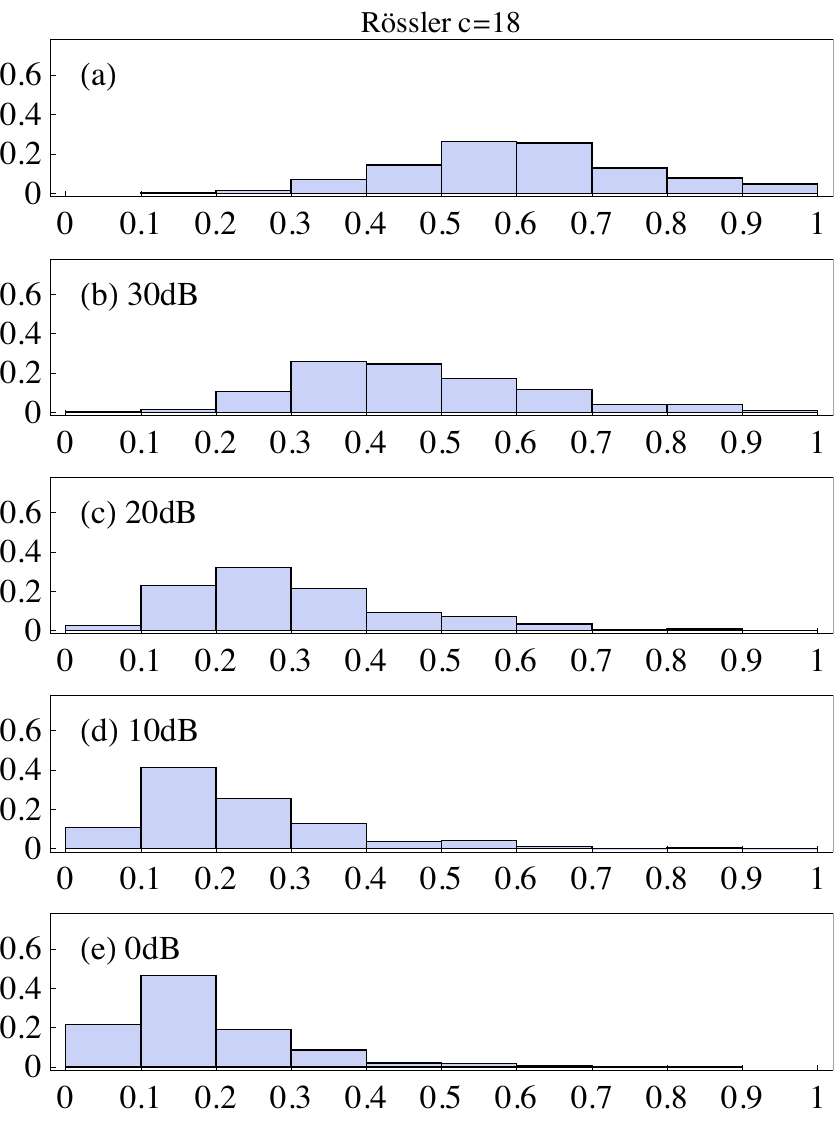}
\parbox{8cm}{\caption{\label{fig:ccnoise}
Clustering coefficient distribution for the chaotic R\"{o}ssler system with $c=18$ by adding noise of different levels:
(a) no noise,
(b) $30$dB white Gaussian noise,
(c) $20$dB white Gaussian noise,
(d) $10$dB white Gaussian noise,
(e) $0$dB white Gaussian noise.
}}
 \end{figure}

Figure~\ref{fig:ccnoise} shows the clustering coefficient distribution for the same set of data with different noise levels. The phase space network exhibits a large clustering coefficient for the chaotic R\"{o}ssler system in the absence of noise. We find the bell shape distribution for clustering coefficients with the maximum values around 0.5. Adding noise results in the left to right shift in the distribution which means the noise tends to destroy the clustering structures in the network.

In the above discussion, we examined the  impact of noise on the network properties from a global view. The existence of multiscale descriptions given by network theory will provide us with more detailed information by measuring the local vertex properties as well. In the next section we will demonstrate that this method can also be usefully applied to experimental time series data. In this case the experimental data comes from a ``toy'' physical experiment. Nonetheless, the methods could also be applied to any system of particular interest.

\section{\label{sec4:app}Analysis of time series from clarinet data}
As a real-world example, we apply the network method to experimental data of sustained tones voiced on a standard $B\flat$ clarinet over most of the dynamic range of the instrument --- from E$_3$ to F$_6$ in standard scientific notation.


A similar application was initially described in Ref.~\onlinecite{Donner:Recurrence:2011}.
The clarinet tones are produced by self-sustained oscillations coupling with nonlinear interactions between the flow and the mechanical response of the reed. By making use of the same complex network method for several sets of time series data from the clarinet, the preliminary result which has been reached is that as the data belongs to superfamilies which are most similar to chaotic or hyperchaotic dynamics, the pure sustained clarinet tones could be characteristically aperiodic, and consistent with chaotic or hyperchaotic dynamics. However, the dynamics in the clarinet tones which we observed could be affected by either the performer's skills or the quality of the clarinet~\cite{Donner:Recurrence:2011}.


In this paper, to extend the same analysis method on clarinet tones, we make use of another set of clarinet data from the stationary period of 23 distinct notes individually recorded at a sampling frequency of 44.1kHz. After being smoothed by an appropriate window width, each data was down-sampled to a level with approximately 25 samples per cycle so that it covers more or less the same number of orbits in the phase space. The time delay was chosen to be the first minimum of mutual information (typically between 2 and 8) and the embedding dimension was 10 which we verified to be significantly larger than the estimated correlation dimension of all data (between 1 and 3). By linking each vertex with its 4 nearest neighbors, the phase space networks have been constructed from the embedded time series.

\begin{table*}[htbp]
\caption{\label{tbl:motifclarinet}
Motif ranking for experimental data of sustained tones voiced on a standard $B\flat$ clarinet}
\begin{center}
\begin{tabular}[c]{|c|c|c|c|c|c|}\hline
Notes &			Motif ranking &		Mean bc & mean cc & average path &	 diameter \\[0.5ex]\hline\hline
E$_3$ &			$ABDCEF$ &		4.070686e+005 &	0.39315 &   41.7109	 &86\\\hline	
F$_3$ &			$ABDCEF$ &		4.647970e+005 &	0.35716 &   47.4843	 &96\\\hline	
G$_3$ &			$ADBCEF$ &		3.598008e+005 &	0.31339 &   36.9837	 &75\\\hline	
A$_3$ &			$ABDCEF$ &		7.344845e+005 &	0.39794 &	74.0342	 &154\\\hline
B$_3$ &			$ABDCEF$ &		1.006649e+008 &	0.37172 &	71.5472	 &144\\\hline	
C$_4$ &			$ABDCEF$ &		1.337342e+019 &	0.46234 &	83.5449	 &169\\\hline	
D$_4$ &			$ABDCEF$ &		5.555720e+005 &	0.40947 &   56.5618	 &113\\\hline	
E$_4$ &			$ABDCEF$ &		4.973566e+015 &	0.39809 &   98.4771	 &200\\\hline	
F$_4$ &			$ABDCFE$ &		4.994924e+008 &	0.41626 &   61.9913	 &125\\\hline	
G$_4$ &			$ABDCEF$ &		4.866596e+007 &	0.37867 &   74.1277	 &150\\\hline	
A$_4$ &			$ABDCEF$ &		6.691653e+005 &	0.37456 &	65.0096	 &131\\\hline	
B$_4$ &			$ABDCEF$ &		6.273676e+016 &	0.41829 &   95.0875	 &192\\\hline	
C$_5$ &			$ABDCEF$ &		6.491825e+005 &	0.32339 &   65.0732	 &132\\\hline	
D$_5$ &			$ABDCEF$ &		5.083065e+019 &	0.40712 &   119.3084 	 &240\\\hline	
E$_5$ &			$ABDCEF$ &		5.779329e+005 &	0.38586 &   58.7792	 &119\\\hline	
F$_5$ &			$ABDCFE$ &		7.343508e+012 &	0.44612 &   79.1254	 &160\\\hline	
G$_5$ & 			$ABDCFE$ &		1.531236e+007 &	0.4283  &	57.275	 &114\\\hline	
A$_5$ &			$ABDCFE$ &		2.127506e+007 &	0.4472  &	67.9874	 &137\\\hline	
B$_5$ &			$ABDCEF$ &		1.074712e+006 &	0.3806  &	86.8553	 &175\\\hline	
C$_6$ &			$ABCDFE$ &		1.394829e+030 &	0.56559 &   100.6292 	 &205\\\hline	
D$_6$ &			$ABDCEF$ &		2.350170e+020 &	0.41137 &   94.4897	 &191\\\hline	
E$_6$ &			$ABDCFE$ &		1.877442e+019 &	0.50417 &   107.3988	 &217\\\hline	
F$_6$ &			$ABDCEF$ &		1.663571e+013 &	0.31788 &   114.8136	 &232\\\hline	
\end{tabular}
\end{center}
\end{table*}

A single motif ranking, $ABDCEF$ dominates. In all but two cases, $ABCDFE$ or $ABDCEF$ is the appropriate motif ranking. The two exceptions are C$_6$ and G$_3$. The note C$_6$ corresponds to one at the very top of the range of the second register (the
so-called clarion register) and is obtained with the very shortest possible air column length, and the fundamental harmonic removed. This note therefore represents an extremity of the range of the instrument. Of course, higher notes are possible (into the altissimo register --- such notes are obtained by removing still further harmonics), however, these are produced with much less regular fingering patterns and require tighter control of the reed by the musician --- to ensure that those lower harmonics are correctly removed. As far as we know, the note G$_3$ is not peculiar --- it is one of the lowest notes of the bottom (chalumeau register), but not the lowest.  The predominant motif ranking which we observe corresponds to that for chaotic and periodic flows with moderate noise. The large path length and diameter are consistent with periodic flows exhibiting small to moderate noise. The moderate clustering coefficient and large betweenness are similarly consistent with a large degree of spatial heterogeneity in the underlying attractor. In fact, more local information is contained in the local scale rather than the average betweenness properties.

While we have not been able to provide a definitive answer to the question of whether clarinet tones are ``chaotic'', the evidence we have obtained is certainly not inconsistent with that proposition. Of course, the issue of whether the intonation could be generated by a noisy periodic orbit remains unaddressed. To resolve this we will need to resort to surrogate data methods
~\cite{Small:Applying:2002,Small:Detecting:2003,Small:Surrogate:2001}.
Nonetheless, it is encouraging that consistent results have been obtained across the full spectrum of experimental data.

\section{\label{sec5:con}Conclusions}

The purpose of this paper has been to provide a more thorough investigation of the characteristic properties of complex networks obtained from times series, specifically using the phase space method of Xu {\em et al.}~\cite{Xu:Superfamily:2008}.
To do so we have not only provided a more detailed exploration of the effect of noise contamination on the motif superfamily methods, but we have also extended the analysis of corresponding networks to include micro- and meso- scale properties as well as the usual global summary statistics. This approach has provided new insights as we have been able to connect dynamical properties of the underlying system to variations in complex network properties of individual nodes. In particular, average path length, global clustering coefficient and degree distribution defined in the global network scale can be used to distinguish systems of different dynamical types. Local degree, clustering coefficients and betweenness centrality contain information on the spatial fillings of the phase space and are close to invariant objects such as UPOs. The specific vertex properties can provide more detailed information about the local attractor geometry in the phase space.

We demonstrated that the method can also be usefully applied to real experimental data without being overwhelmed by the effect of noise. In this case, the natural, but still rather obvious conclusion is that clarinet sound production is consistent with either pseudoperiodic chaos or a noise contaminated periodic orbit. Nonetheless, this is a useful conclusion as methods exist to distinguish between these two cases, and the application of such methods with the measures described here as test statistics will be straightforward~\cite{Small:Applying:2002,Small:Detecting:2003,Small:Surrogate:2001}.

\begin{acknowledgments}
This work was supported by a Hong Kong Polytechnic University direct allocation (G-YG35). X.-K.X. was supported by the PolyU Postdoctoral Fellowships Scheme (G-YX4A) and the Research Grants Council of Hong Kong (BQ19H). J.Z. and X.-K.X. also acknowledges the National Natural Science Foundation of China (61004104, 61104143).
\end{acknowledgments}
\bibliography{ref_3}
\end{document}